\documentclass[final,5p,times,twocolumn]{elsarticle}

\usepackage{amssymb}

\usepackage{color}
\usepackage{amsmath}
\usepackage{graphicx}
\usepackage[dvipsnames]{xcolor}
\usepackage{bm}
\usepackage{epstopdf}
\usepackage[colorlinks,citecolor=blue]{hyperref} 

\journal{Physics Letters A}

\begin{document}

\begin{frontmatter}



\title{Comment on ``Adaptive modification of the delayed feedback control algorithm with a continuously varying time delay" [Phys. Lett. A 375 (44) (2011) 3866–3871]}


\author{Viktor Novi\v{c}enko}

\address{Faculty of Physics, Vilnius University, Saul\.{e}tekio ave. 3, LT-10257 Vilnius, Lithuania}

\begin{abstract}
In the paper~\cite{Pyragas2011} authors propose	a modification of the conventional delayed feedback control algorithm, where time-delay is varied continuously to minimize the power of control force. Minimization is realized via gradient-descent method. However, the derivation of the gradient with respect to time-delay is not accurate. In particular, a scalar factor is omitted. The absolute value of the scalar factor is not crucial, as it only changes the speed of the gradient method. On the other hand, the factor's sign changes the gradient direction, therefore for negative value of the multiplier the gradient-decent becomes gradient-ascent method and fail power minimization. Here the accurate derivation of the gradient is presented. We obtain an analytical expression for the missing factor and show an example of the Lorenz system where the negative factor occurs. We also discuss a relation between the negativeness of the factor and the odd number limitation theorem.
\end{abstract}



\end{frontmatter}



Following~\cite{Pyragas2011}, let us assume that we have single-input-single-output dynamical system ${\dot{\mathbf{X}}=\mathbf{F}\left(\mathbf{X}, u_{\mathrm{in}} \right)}$ where $u_{\mathrm{in}}(t)$ represents scalar input signal, and a scalar output signal $s(t)=g \left[ \mathbf{X}(t) \right]$ is available. We tend to stabilize unstable periodic orbit $\bm{\xi}(t+T)= \bm{\xi}(t)$, which is a solution of the dynamical system for $u_{\mathrm{in}}=0$. To do that, the force of the form of the delayed feedback control $u_{\mathrm{in}}(t)=K[s(t)-s(t-\tau)]$ is used. The time-delay $\tau(t)$ is assumed to be slowly changing (in comparison to the period $T$) dynamical variable which tends to minimize a potential $\Delta(t)$:
\begin{equation} \label{eq:pot}
\Delta(t)=\int\limits_{t_0}^{t} \mathrm{e}^{-\nu(t-t^\prime)}\left[ s(t^\prime)-s(t^\prime-\tau(t))\right]^2 \mathrm{d}t^\prime,
\end{equation}
defined as an exponentially weighted average of square of the control force. Here $t_0$ is the time moment when the controller is turned on, and $\nu^{-1}>T$ represents the characteristic width of the averaging window. The time-delay is varying slowly with respect to negative gradient $\dot{\tau}=-\beta G(t)$, here the positive parameter $\beta^{-1}>\nu^{-1}$ represents speed of the gradient-descent method, and the dynamical variable $G(t)=\mathrm{d}\Delta/\mathrm{d} \tau$ is the gradient of the potential ~(\ref{eq:pot}) with respect to the time-delay. The inaccuracy of~\cite{Pyragas2011} occurs in the calculation of $\mathrm{d}\Delta/\mathrm{d} \tau$, in particular the authors assume that
\begin{equation} \label{eq:sder}
\frac{\mathrm{d}}{\mathrm{d} \tau} s(t)=0 \quad \mathrm{and} \quad  \frac{\mathrm{d}}{\mathrm{d} \tau}s(t-\tau)=-\dot{s}(t-\tau),
\end{equation}
and thus obtain differential equation
\begin{equation} \label{eq:grad}
\dot{G}=2[s(t)-s(t-\tau)]\dot{s}(t-\tau)-\nu G
\end{equation}
for the gradient. We claim that~(\ref{eq:sder}) is not derived correctly and as a consequence~(\ref{eq:grad}) does not represent the gradient.

In general, the accurate derivation of $G(t)$ is a difficult task. However one can do it by assuming that a quantity $(\tau-T)$ is small and perform expansion of $G(t)$ with respect to $(\tau-T)$. For the time-delay $\tau$ being close to $T$, the dynamical system under delayed feedback control performs periodic behavior with the period (see Refs.~\cite{just98,nov12})
\begin{equation} \label{eq:theta}
\Theta(\tau)=T+\left[1-\alpha(KC)\right](\tau-T)+\mathcal{O}\left[(\tau-T)^2\right],
\end{equation}
where function $\alpha(x)=(1+x)^{-1}$ and the constant $C$ is the following integral~\cite{nov12}:
\begin{equation} \label{eq:c}
C=\int\limits_0^T \left\lbrace \mathbf{z}^T(t)\cdot D_2 \mathbf{F}\left( \bm{\xi}(t),0 \right) \right\rbrace \left\lbrace \left[ \nabla g\left(\bm{\xi}(t)\right) \right]^T \cdot \dot{\bm{\xi}}(t) \right\rbrace \mathrm{d}t.
\end{equation}
Here superscript $(\,)^T$ denotes transposition, $D_2$ represents the derivative of the vector field with respect to the second argument and $\mathbf{z}(t+T)=\mathbf{z}(t)$ is a phase response curve of the periodic orbit $\bm{\xi}(t)$ of the control-free dynamical system. 

Using~(\ref{eq:theta}) one can expand the output signal in terms of small parameter $(\tau-T)$. The zeroth-order approximation gives
\begin{equation} \label{eq:s_zero}
s(t)=g\left[\bm{\xi}\left(t\frac{T}{\Theta(\tau)}\right)\right]+\mathcal{O}(\tau-T).
\end{equation}
The crucial moment here is that in order to preserve periodicity of $s(t)$ with the period $\Theta$ on long time interval $t \sim 1/(\tau-T)$ one should put the argument $tT/\Theta$ instead of simple $t$ into $\bm{\xi}(\cdot)$. From~(\ref{eq:s_zero}) now is clearly seen that~(\ref{eq:sder}) does not hold. In particular the first derivative in Eq.~(\ref{eq:sder}) gives:
\begin{equation} \label{eq:sder1}
\frac{\mathrm{d}}{\mathrm{d} \tau} s(t)= \left\lbrace \nabla g \left[ \bm{\xi} \left( t\frac{T}{\Theta} \right) \right] \right\rbrace^T \cdot \dot{\bm{\xi}}\left( t\frac{T}{\Theta} \right)t \frac{\mathrm{d} (T/\Theta)}{\mathrm{d} \tau}.
\end{equation}
The fraction $T/\Theta$ up to second order term gives
\begin{equation} 
\frac{T}{\Theta}=1+\frac{\alpha-1}{T}(\tau-T).
\label{eq:fra}
\end{equation}
Using periodicity feature, the delayed signal can be written as
\begin{equation}
\label{eq:del}
\begin{split}
& s(t-\tau)=s\left(t+\Theta-\tau\right)=s(t-(\tau-T)\alpha)  \\
& =g\left[ \bm{\xi} \left(t\frac{T}{\Theta}-(\tau-T)\alpha \frac{T}{\Theta} \right) \right].
\end{split}
\end{equation}
Using~(\ref{eq:fra}), the term $(\tau-T)\alpha T/\Theta$ simplifies to $(\tau-T)\alpha$, since the second order term $(\tau-T)^2$ might be omitted. Thus we have
\begin{equation} 
s(t-\tau)=g\left[ \bm{\xi} \left(t\frac{T}{\Theta}-(\tau-T)\alpha  \right) \right].
\label{eq:del1}
\end{equation}
Now we can calculate the second derivative in Eq.~(\ref{eq:sder}):
\begin{equation}
\label{eq:sder2}
\begin{split}
\frac{\mathrm{d}}{\mathrm{d} \tau} s(t-\tau) =& \left\lbrace \nabla g \left[ \bm{\xi} \left( t\frac{T}{\Theta}-(\tau-T)\alpha \right) \right] \right\rbrace^T \\
& \cdot \dot{\bm{\xi}}\left( t\frac{T}{\Theta} -(\tau-T)\alpha \right)\left[ t \frac{\mathrm{d} (T/\Theta)}{\mathrm{d} \tau} -\alpha \right].
\end{split}
\end{equation}
By subtracting~(\ref{eq:sder2}) from~(\ref{eq:sder1}) and collecting only the zeroth-order terms (one can see that after subtraction the terms $(\tau-T)$ in the right hand side of~(\ref{eq:sder2}) might be ignored without loss of the accuracy) we finally obtain
\begin{equation} \label{eq:smins}
\frac{\mathrm{d}}{\mathrm{d} \tau} \left[ s(t)-s(t-\tau) \right]=\alpha \left\lbrace \nabla g \left[ \bm{\xi} \left( t\frac{T}{\Theta} \right) \right] \right\rbrace^T \cdot \dot{\bm{\xi}}\left( t\frac{T}{\Theta} \right).
\end{equation}
Despite complexity of the right hand side of last expression, it can be obtained as a time derivative of the output signal multiplied by $\alpha$:
\begin{equation} \label{eq:smins1}
\frac{\mathrm{d}}{\mathrm{d} \tau} \left[ s(t)-s(t-\tau) \right]=\alpha \dot{s}(t).
\end{equation}
By comparing~(\ref{eq:smins1}) with the result obtained in Ref.~\cite{Pyragas2011}, one can see two differences: the omitted time-delay in the derivative and the additional factor $\alpha$. The first difference is not essential, since both $\dot{s}(t)$ and $\dot{s}(t-\tau)$ well approximate the left hand side of~(\ref{eq:smins1}) up to zeroth-order accuracy. Here we use $\dot{s}(t)$ only for the sake of simplicity. The second difference may be crucial. The magnitude of the factor $\alpha$ is not so critical, since it only re-scales the speed of gradient-descent method $\beta$. On the other hand, the sign of $\alpha$ is essential as it may change direction of the gradient and as a consequence the control method fail the stabilization.

The positive factor $\alpha$ implies correct direction of the gradient. For example, $\alpha>0$ is in the case of the stabilization of the periodic orbit in R\"{o}ssler and Mackey--Glass systems analyzed in Ref.~\cite{Pyragas2011}. However, the case of $\alpha<0$ gives incorrect sign of the gradient $G(t)$ governed by Eq.~(\ref{eq:grad}). The negative $\alpha$ emerges in the situations, where control free periodic orbit has an odd number of Floquet multipliers larger than one. Such a periodic orbit fall under odd number limitation theorem~\cite{hoo12} and can be stabilized by delayed feedback control only with the the control gain $K$ values such that $\alpha(KC)<0$. This is exactly the case for Lonenz system demonstrated further in the text.

\begin{figure} [h!]
\centering\includegraphics[width=0.95\columnwidth]{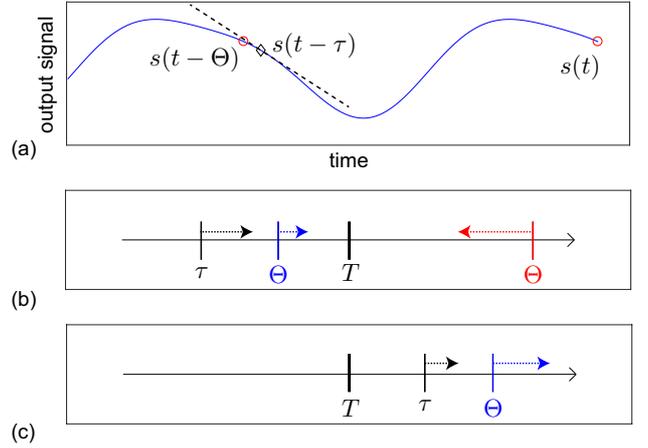}
\caption{\label{fig1} Schematic analysis of the gradient-descent method presented in~Ref.~\cite{Pyragas2011}. (a)  Periodic output signal $s(t)$. (b)  Numerical axis and the motion of the quantities $\tau$ and $\Theta$ for the positive $\alpha$. (c) The same as in (b) only for the negative $\alpha$.}
\end{figure}

The reason why the gradient-descent method presented in~\cite{Pyragas2011} works for $\alpha>0$, but does not work for $\alpha<0$ can be illustrated by the following schematic analysis. Let's say that the periodic output signal $s(t)=s(t+\Theta)$ has a form depicted in Fig.~\ref{fig1}(a) and $\Theta>\tau$. Red circles indicate values of $s(t)$ and $s(t-\Theta)$, while black diamond shows $s(t-\tau)$. The difference ${[s(t)-s(t-\tau)]}$ is positive and the derivative $\dot{s}(t-\tau)$, depicted as a tangent dashed line, is negative. Thus the first term in the right hand side of  Eq.~(\ref{eq:grad}) is negative. If $G(t)$ will be defined not as the exponentially weighted average but as an instantaneous gradient, it will be negative too. Therefore $\dot{\tau}=-\beta G$ tends to increase the time-delay or in other words, $\tau$ is moved towards $\Theta$. It is a right strategy for the $\alpha>0$. In Fig.~\ref{fig1}(b) we plot numerical axis and the values of $T$, $\tau$ and $\Theta$ on it for the case of $\alpha>0$. The case of $0<\alpha<1$ plotted by blue color while $\alpha>1$ plotted by red color. Arrows show motion of $\tau$ and $\Theta$. They coincide exactly at the target value $T$. However, for the case of $\alpha<0$, the strategy to shift $\tau$ towards $\Theta$ results to $\tau$ movement away from the target value $T$, as it is seen from Fig.~\ref{fig1}(c).

To sum it up, let us write out revised version of equations (10)--(13) in Ref.~\citep{Pyragas2011}:
\begin{subequations}
\label{eq:full}
\begin{align}
\dot{\mathbf{X}} &= \mathbf{F} \left(\mathbf{X}, K\left[ s(t)-s(t-\tau) \right] \right), \label{eq:full_1} \\
\dot{\tau} &= -\beta G, \label{eq:full_2} \\
\dot{G} &= 2 \gamma \alpha \left[s(t)-s(t-\tau) \right]\left[s(t)-u \right]-\nu G, \label{eq:full_3} \\
\dot{u} &= \gamma \left[s(t)-u \right],
\end{align}
\end{subequations}
where the parameter's values $\gamma^{-1}<T<\nu^{-1}<\beta^{-1}$ must be keep in mind. Typically $\alpha$ is not known. The magnitude of $\alpha$ might be ignored by putting $\alpha=\mathrm{sgn}[\alpha]$, as it just rescale the speed of gradient-descend $\beta \rightarrow \beta |\alpha|$. However, a choice for the right sign should be done by a trial and error method.

The expression~(\ref{eq:smins1}) gives the zeroth-order term while $\left[ s(t)-s(t-\tau) \right]$ is the first-order term, thus Eq.~(\ref{eq:full_3}) approximates the gradient $\mathrm{d}\Delta/\mathrm{d}\tau$ up to the first-order, i.e. ${G(t)\propto \mathcal{O}(\tau-T)}$.

\begin{figure}[h!]
	\centering\includegraphics{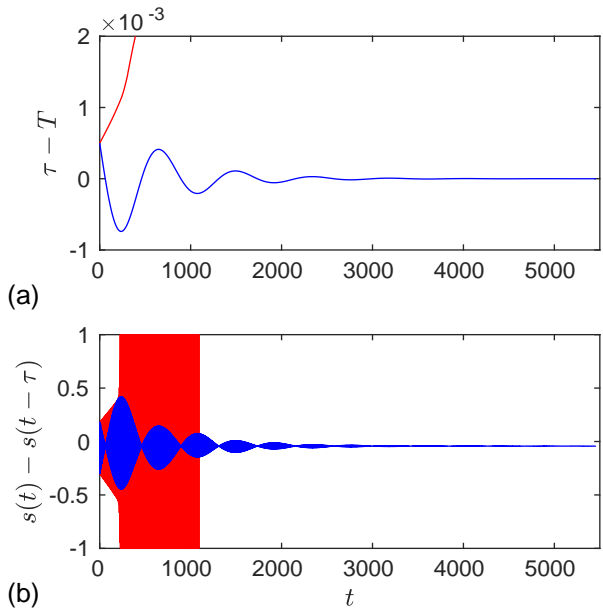}
	\caption{\label{fig2} Application of adaptive delay feedback algorithm for Lorenz system (\ref{lorenz}). Red lines represent method described in Ref.~\cite{Pyragas2011}, blue lines -- corrected version (\ref{eq:full}).  (a) The difference between state dependent delay $\tau(t)$ and the period of unstable orbit $T$. (b) Applied control force. The integration was performed by using standard MATLAB function \textit{ddesd}.  Parameters: $K=1$, $\gamma=50/T$, $\nu=1/(200T)$, $\beta = 10^{-9}/T$.}
\end{figure}

Further, we will provide an application of adaptive delayed feedback algorithm for the period-one unstable periodic orbit stabilization for the Lorenz system. The orbit has a singe Floquet multiplier larger that one, thus it can be stabilized only with $\alpha<0$. The state vector $\mathbf{X}(t)=\left[ x_1(t), x_2(t), x_3(t) \right]^T$ evolves according to the vector field:
%
\begin{equation}
\mathbf{F}\left( \mathbf{X},u_{\mathrm{in}} \right)= \left[\begin{array}{c}
10(x_2-x_1)\\
x_1(28-x_3)-x_2+u_{\mathrm{in}}\\
x_1x_2-8/3x_3\\
\end{array}\right].\label{lorenz}
\end{equation}
The scalar output signal has the form of $s(t) = x_2/2-x_1$, adopted from Ref.~\cite{novicenko13}. For a given model, the period of unstable orbit is $T \approx 1.558652$ and the constant $C=-1.285$ (see~\cite{nov12,novicenko13} for a more detail on how the constant $C$ can be calculated). We set $K=1$, therefore $\alpha \approx -3.5$. In Fig.~\ref{fig2}(a) the evolution of difference between the state dependent delay $\tau(t)$ and period $T$ is demonstrated, where red line represents method reported in~\cite{Pyragas2011}, blue line -- corrected version (\ref{eq:full}). As predicted by our reasoning, the red curve moves away from correct value of $T$, while blue slowly approaches it. The Fig.~\ref{fig2}(b) confirms that (\ref{eq:full}) gives correct delay value as the applied control force tends to zero (blue line), while for~\cite{Pyragas2011} method (red line) the  control force rises to high values.

%





\bibliography{references}

\end{document}